\documentclass[aps,prc,preprint,superscriptaddress]{revtex4}

\usepackage{graphicx}
\usepackage{bm}
\usepackage{wrapfig}
\usepackage{amsmath,amssymb}
\usepackage{color}

\newcommand{\bra}[1]{\langle {#1} |}
\newcommand{\ket}[1]{| {#1} \rangle}

\def\vect#1{{\mbox{\boldmath $#1$}}}

\begin{document}


\title{Nuclear incompressibility parameters evaluated from\\
isoscalar giant monopole resonance of $N=Z$, $A=100, 132$ nuclide and Sn isotopes}


\author{Shuichiro~Ebata}
\affiliation{Faculty of Science, Hokkaido University, Sapporo, 060-0810, Japan}


\date{\today}

\begin{abstract}
The isoscaler giant monopole resonances (ISGMR) are computed 
using the canonical-basis time-dependent Hartree-Fock-Bogoliubov theory 
(Cb-TDHFB) with five kinds of Skyrme parameter sets 
(SGII, SkM$^*$, SLy4, SkT3 and SkI3). 
To extract the nuclear matter property from finite system, 
ISGMRs of $N$=$Z$ ($Z$=20 - 50), isobar even-even nuclide for $A$=100,
132 and Sn isotopes are analysed systematically. 
The magnitude relation of nuclear incompressibility-parameter ($K_\infty$) 
among Skyrme parameter sets, can be corresponded to 
the peak positions of GMR in spherical isotopes over $A$=80. 
The parameters ($K_{\rm surf}, K_\tau$ and $K_{\rm Coul}$) 
which appear in expansion of the finite nucleus incompressibility $K_A$, 
are determined for each Skyrme parameter. 
From the comparison experimental data whole mass region 
and the present results, 
they indicate that the isospin dependent term $K_\tau$ 
is filtered as -305$\pm$10 MeV. 
The incompressibility parameters of {\it infinite} system 
corresponding to our results 
are $K_\tau^\infty$=-340$\pm$35, $K_\infty$=225$\pm$11, 
and $K_{\rm sym}$=-138$\pm$18 MeV.
\end{abstract}


\maketitle

\section{Introduction}
To extract an equation of state (EOS) for nuclear matter from finite nuclear system, 
is one of most important task given to nuclear physics. 
The EOS is a very important topic to connect the nuclear physics to  
the astrophysical objects such as nucleosynthesis, neutron star, and so on. 
The EOS is often expressed in the expansion around symmetric matter, as follows. 
\begin{eqnarray}
\frac{E}{A}[\rho,\delta]&=&{\cal E}[\rho,\delta=0] + {\cal S}[\rho,\delta] + {\cal O}[\delta^4], \nonumber \\
{\cal E}[\rho]&\equiv& {\cal E}_0 + \frac{K_\infty}{2}\varrho^2 + \frac{Q_0}{6}\varrho^3 + \cdots, \nonumber \\
{\cal S}[\rho,\delta]&\equiv& \left( J + L\varrho + \frac{K_{\rm sym}}{2}\varrho^2 + \cdots \right)\delta^2, 
\label{EoS}
\end{eqnarray}
where $\delta=(\rho_n-\rho_p)/\rho$ is an asymmetric parameter 
which separates a symmetric matter (SM) EOS ${\cal E}$ 
and symmetry energy ${\cal S}$. 
They are expressed in $\varrho=(\rho-\rho_0)/3\rho_0$, which is 
an expansion around the nuclear saturation density $\rho_0$ 
at which the pressure of nuclear matter is zero. 

When the EOS is expanded as Eq.(\ref{EoS}), there appear 
characteristic parameters: the binding energy per nucleon ${\cal E}_0$, 
the incompressibility of SM $K_\infty$, 
the skewness parameter of SM $Q_0$, the symmetry energy $J$, 
the slope parameter $L$ and the symmetry incompressibility $K_{\rm sym}$, and so on. 
The ${\cal E}_0$ and $J$ are respectively equal to 
the ${\cal E}$ and ${\cal S}$ at $\rho_0$. 
The other parameters can be obtained 
from the density derivation of ${\cal E}$ and ${\cal S}$ as follows.
\begin{eqnarray}
&&K_\infty = 9 \rho_0^2 \left. \frac{\partial^2 {\cal E}}{\partial \rho^2} \right|_{\rho=\rho_0},\\[2mm]
&&Q_0 = 27 \rho_0^3 \left. \frac{\partial^3 {\cal E}}{\partial \rho^3} \right|_{\rho=\rho_0},
\end{eqnarray}
\begin{eqnarray}
&&L = 3 \rho_0 \left. \frac{\partial {\cal S}}{\partial \rho} \right|_{\rho=\rho_0},\\[2mm]
&&K_{\rm sym} = 9 \rho_0^2 \left. \frac{\partial^2 {\cal S}}{\partial \rho^2} \right|_{\rho=\rho_0}. 
\end{eqnarray}
Thus the EOS parameters can be easily obtained 
at when the energy density functional (EDF) $E/A [\rho]$ is chosen. 
Nucleus is a too much small system to extrapolate 
directly to the infinite nuclear matter \cite{Bla80}. 
Therefore, we evaluate the EOS with 
the help of effective interaction (Skyrme, Gogny, relativistic mean field) 
\cite{SY07,SR07,PC09,Che10,Du12,Br13,CP14,Du14}. 
However, it is not easy to determine acceptable values 
in nuclear structure and in astrophysics, at same time. 
Currently also still, a study of the EOS using effective interactions 
has been progressed theoretically 
and also experimentally \cite{Li07,PGF12,PGF13,Tam11}. 
The uncertainties of ${\cal E}_0$ and $J$ are expected to be small: 
${\cal E}_0 \approx$ 16 MeV which has appeared in 
the Bethe-Weiz$\ddot{\rm a}$cker mass formula, 
and $J$=32$\pm$3 MeV \cite{SR07}. 
The $K_\infty$ does not have so large uncertainty due to the consistency among 
experiments and theoretical prediction: $K_\infty$=230$\pm$30 MeV \cite{Du12}.  
The slope parameter $L$ which will be strongly related to nuclear dipole mode, 
has been well studied from the many points of view, for instance 
the relation among the neutron-skin thickness \cite{War09,Car10,Roca11,RN10, Tam11}, 
pygmy dipole resonance \cite{Car10}, 
giant dipole resonance (GDR) \cite{Tri08} and polarizability \cite{Tam11,RN10,War14}, 
although it has been yet floated: $L$=58$\pm$18 MeV \cite{Che10,Du14}. 
These untiring studies narrow downs the range of the EOS parameters, 
however the $K_{\rm sym}$ especially has a large range of values. 
Basically, it is difficult to connect between EOS parameters and experimental values 
directly, therefore we usually take the procedure: 
to search an EDF to reproduce experiments and 
then to extract the EOS parameters from the EDF. 
In a present work, we evaluate incompressibility parameters 
from the isoscalar giant monopole resonance (ISGMR) of finite nuclear system. 
In order to estimate the parameters independently of the speciality of each nucleus, 
we use the finite nuclear incompressibility $K_A$-expansion to analyse them, 
although we can compare the ISGMRs in theory and in experiments. 

If the energy of ISGMR $E_{\rm GMR}$ is represented 
in the root mean square radius of the nucleus and 
the GMR can be regarded as a single phonon mode, $E_{\rm GMR}$ is written in 
\begin{eqnarray}
E_{\rm GMR}=\sqrt{\frac{\hbar^2 K_A}{m \langle r^2 \rangle}}, 
\label{EGMR}
\end{eqnarray}
where $K_A$ is defined as an incompressibility of finite nuclear system \cite{Bla80}. 
Although the $K_A$ can not directly equal to $K_\infty$ at a limit of $A\to\infty$, 
it will bring the relation between the parameters of finite and infinite system.
The $K_A$ is expanded around $K_\infty$ as 
\begin{eqnarray}
K_A\! =\! K_{\infty}\! +\! K_{\rm surf}A^{-1/3} + K_\tau \!\! \left(\! \frac{N\!-\!Z}{A} \right)^2 
\!\!+ K_{\rm Coul} \frac{Z^2}{A^{4/3}}. \nonumber \\
\label{KAex}
\end{eqnarray}
When incompressibility parameters are extracted from the experiments, 
the expansion (\ref{KAex}) is often used \cite{Li07,PGF12,PGF13,SSM14}. 
The surface and Coulomb term ($K_{\rm surf}$, $K_{\rm Coul}$) 
are estimated by infinite EOS parameters in previous studies, 
which will be confirmed how available in this work. 
The isospin term $K_\tau$ has not been clear, which 
is a most important key to know the information 
of nuclear matter incompressibility using Eq.(\ref{KAex}). 

This paper is organized as follows. 
In Sec.\ref{sec:method}, we present the methods to calculate ISM mode and 
to evaluate the centroid energy of ISGMR $\bar{E}_{\rm GMR}$. 
In Sec.\ref{sec:Result}, 
at first, we show partially the strength functions of $N=Z$ nuclide 
to check the behaviour of them including the results of deformed nuclei.
After that, we extract the EOS parameters of finite nuclear system 
by the chi-square fitting with Eq.(\ref{KAex}), 
from the $K_A$ evaluated with Eq.(\ref{EGMR}) and $\bar{E}_{\rm GMR}$.
In particular, we investigate $N=Z$ nuclide for $K_{\rm surf}$ 
and $K_{\rm Coul}$, and isobar nuclide for $K_\tau$.
Furthermore, to confirm the expansion itself and with our coefficients, 
we compare them with the experimental $E_{\rm GMR}$ 
and calculated $\bar{E}_{\rm GMR}$ for Sn isotopes. 
In Sec.\ref{sec:discuss}, 
we also compare the experimental measurements 
for a whole mass region with $A$=24$-$238, 
in order to narrow down the candidates of effective interaction. 
And we mention the mass dependence of $E_{\rm GMR}$ in experiments. 
The present work is summarized in Sec.\ref{sec:con}.
\label{sec:introduction}

\section{Formulation}
To access the GMR, 
we apply the Canonical-basis time-dependent Hartree-Fock-Bogoliubov (Cb-TDHFB) 
theory \cite{EN10} in three-dimensional (3D) coordinate space 
which can be successfully applied to the study of the dipole \cite{EN10, EN14} and 
quadrupole \cite{SL13,EK15} modes of many isotopes, systematically. 
The Cb-TDHFB can describe the dynamical effects 
of pairing correlation in fully self-consistently. 
The Cb-TDHFB equations are derived from
the full TDHFB equation represented in the canonical basis 
$\{ \phi_l(t), \phi_{\bar l}(t)\}$ which diagonalize a density matrix, 
and by assuming the diagonal form of pairing functional. 
The Cb-TDHFB equations compose the time-evolution equations 
for the canonical pair $\{ \phi_l(t),\phi_{\bar l}(t) \}$, 
its occupation probability $\rho_l(t)$ and pair probability $\kappa_l(t)$,  
\begin{eqnarray}
\label{Cb-TDHFB}
&i&\frac{\partial}{\partial t} \ket{\phi_l(t)} =
(h(t)-\eta_l(t))\ket{\phi_l(t)},\\ 
&i&\frac{d}{dt}\rho_l(t) =
\kappa_l(t) \Delta_l^{\ast}(t)
-\kappa_l^{\ast}(t) \Delta_l(t) , \nonumber\\
&i&\frac{d}{dt}\kappa_l(t) =
\left(
\eta_l(t)+\eta_{\bar l}(t)
\right) \kappa_l(t) + \Delta_l(t) \left( 2\rho_l(t) -1 \right) ,\nonumber
\end{eqnarray}
where the phase of canonical basis is chosen as 
$\eta_l(t)\equiv\bra{\phi_l(t)}h(t)\ket{\phi_l(t)}$, 
and the $h(t)$ and $\Delta_l(t)$ are the single-particle Hamiltonian 
and the gap energy, respectively. 

Due to the appearance of deformed ground state in our subjective nuclide, 
we should choose a flexible calculation space. 
We use the 3D Cartesian coordinate-space representation 
for canonical basis, 
$\phi_l(\vect{r},\sigma; t) = \langle \vect{r},\sigma | \phi_l(t) \rangle$ 
with spin $\sigma=\pm 1/2$. 
The condition of calculation space is discretized
in a square mesh of 1.0 fm inside of a sphere of radius 15 fm 
for all nuclide in present work.

To apply our method to systematic investigation, 
we choose the Skyrme EDF to $ph$-channel and the simple pairing functional 
form to $pp$($hh$)-channel: $\Delta_l(t)\equiv\sum_k G_{kl} \kappa_k(t)$ 
where $G_{kl}$ is constant in real-time evolution, as same as Ref. \cite{EN10}. 
Our choice of five Skyrme parameter sets is 
SGII\cite{GS81}, SkM$^*$\cite{Bar82}, SLy4\cite{Cha98}, SkT3\cite{Ton84} 
and SkI3\cite{RF95}. 
The reason to choose them is not only their usefulness, also 
corresponds to the limitation of $K_\infty$=230$\pm$30 MeV 
indicated in Ref.\cite{Du12} 
\label{sec:method}

\subsection{Linear response calculation with Cb-TDHFB}
In order to induce monopole responses, we add a weak instantaneous external field 
$V_{\rm ext}(\bm{r},t)=\xi\hat{F}(\bm r) \delta(t)$ to initial states of the time
evolution. Here the isoscalar monopole operator acting on nucleus 
is given as $\hat{F}_{E0}^{IS} \equiv  r^2 Y_{00}$.  
The amplitude of the external field is so chosen to be a small number
$\xi=2\times10^{-3}$ fm$^{-2}$ to guarantee the linearity.
The strength function $S(E;{\rm E0})$ can be obtained 
through the Fourier transformation of 
${\cal F}(t) \equiv \langle \Psi(t)| \hat{F}_{E0}^{IS} | \Psi(t) \rangle$:  
\begin{eqnarray}
S(E;{\rm E0}) &\equiv& 
\sum_{n} |\langle \Psi_n | \hat{F}_{E0}^{IS} | \Psi_0 \rangle |^2
 \delta(E_n - E) \nonumber \\ 
 &=& 
 \frac{-1}{\pi \xi} {\rm Im} \int_{0}^{\infty} \!\! \left[ {\cal F}(t) - {\cal F}(0) \right] 
 e^{i(E+i\Gamma/2)t} dt,  \nonumber \\
\end{eqnarray}
where $| \Psi_0 \rangle$ and $| \Psi_n \rangle$ are 
the ground and excited states, respectively. 
$\Gamma$ is a smoothing parameter set to 1 MeV for whole nuclide in present. 
\label{sec:Cb-TDHFB}

\subsection{Evaluation of mean energy for GMR}
We need a procedure to compute the mean energy of GR, 
in common among subjective nuclide.
To compute the centroid energy $\bar{E}$ of GR, we use $m_1/m_0$, 
although there are some evaluations ($\sqrt{m_3/m_1}$ or $\sqrt{m_1/m_{-1}}$). 

The $m_1/m_0$ is computed as follows in present work.
\begin{eqnarray}
 \bar{E}_{\rm GMR} \equiv \frac{m_0}{m_1} 
 \simeq \frac{\int_{^>e}^{^<e} dE\ ES(E;{\rm E0})}{\int_{^>e}^{^<e} dE\ S(E;{\rm E0})}, 
\label{bGMR}
\end{eqnarray}
where $^<e$ and $^>e$ are an upper and under cut-off energy respectively. 
They should be decided with more carefully, 
because the $\bar{E}_{\rm GMR}$ is sensitive to them. 
In this work, we decide them as: $^<e = E_{\rm C}+7.5$ 
and $^>e = E_{\rm C}-7.5$ MeV, where $E_{\rm C}=80A^{-1/3}$. 
The empirical formula $E_{\rm C}$ is found in a droplet model \cite{Liu91}. 
Purposely we chose this way, because Eq.(\ref{EGMR}) 
to relate the $K_A$ with nuclear response, 
which is based on the one phonon picture 
in other words the GMR is assumed as one mode.
\label{sec:EGMR}

\section{Result}
We evaluate the mean energy of GMR using Eq.(\ref{bGMR}), and from them 
the finite incompressibility $K_A$ is also evaluated according to Eq.(\ref{EGMR})
with using calculated $\langle r^2 \rangle$ in Table \ref{tab:gs1}. 
To determine the expression parameters of $K_A$ according to Eq.(\ref{KAex}), 
we proceed a following way step by step. 
First, we determine the $K_{\rm surf} and K_{\rm Coul}$ 
using the results of $N=Z$ nuclide in which the isospin term $K_\tau$ 
does not contribute to $K_A$. 
Second, we determine the $K_\tau$ while using the $K_{\rm surf}$ and $K_{\rm Coul}$
fixed in the first step. 
To obtain the $K_\tau$ for each Skyrme interaction, 
we analyse isobar nuclide for both $A=$100 and 132 at same time. 
Lastly, to confirm that the expansion of $K_A$ with determined parameters 
reproduces the results of Sn isotopes, 
we calculate the centroid energies $E_{\rm GMR}$ according to Eq.(\ref{EGMR}) 
with $K_A$ and $\langle r^2 \rangle$ in Table \ref{tab:gs2}, 
and compare them with actually calculated $\bar{E}_{\rm GMR}$s.
\label{sec:Result}

\subsection{$N=Z$}
\begin{figure}[h]
 \begin{center}
  \includegraphics[keepaspectratio,width=5cm, angle=-90]{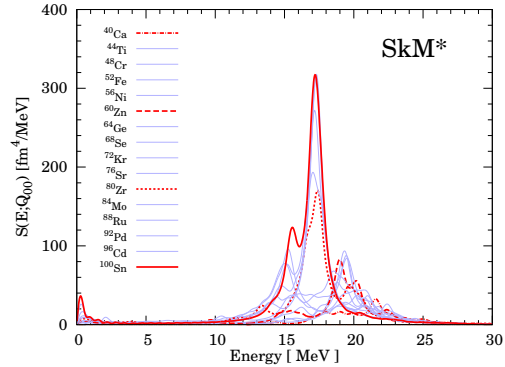}
  \caption{(Colour online) Strength functions of isoscalar monopole vibrational modes 
  of $N=Z$ even-even nuclide from $^{40}$Ca to $^{100}$Sn. 
  }\label{fig:1}
 \end{center}
\end{figure}

ISM strength functions of even-even $N=Z$ nuclide from $^{40}$Ca to $^{100}$Sn 
with SkM$^*$ are shown in Fig.\ref{fig:1}. 
Chain, dashed, doted and thick lines show the results of $Z=$20, 30, 40 and 50, 
respectively. 
We can see the broad strength distribution of $^{40}$Ca in heigh energy 
around 21 MeV. 
The distribution becomes localized and 
its centre shifts to low energy, as mass number increase.
In these strength, split distributions can be seen in thin lines 
which are corresponding to $^{48}$Cr and $^{72}$Kr. 
The split is caused by the coupling monopole with quadrupole excitations 
due to the deformation. 
Typically the quadrupole GR appear in lower energy than GMR, 
thus the $\bar{E}_{\rm GMR}$ of a well deformed nucleus shifts to low energy. 

There are some strengths in vicinity of zero energy, 
which corresponds to numerical spurious mode 
due to the detail of mesh size and of time step. 
They are excluded from the estimation of $\bar{E}_{\rm GMR}$ in Eq.(\ref{bGMR}). 

\begin{figure}[h]
 \begin{center}
  \includegraphics[keepaspectratio,width=5cm, angle=-90]{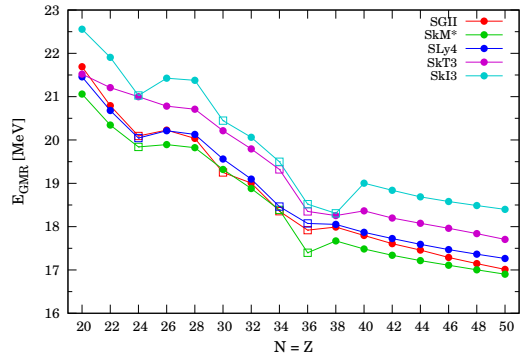}
  \caption{(Colour online) Mean energies of ISGMR for 
  $N=Z$ even-even nuclide, which are computed by Eq.(\ref{bGMR}) 
  using the strengths with five Skyrme parameter sets. 
  }\label{fig:2}
 \end{center}
\end{figure}

Figure \ref{fig:2} shows the $\bar{E}_{\rm GMR}$ for $N=Z$ nuclide with 
five Skyrme parameters. 
Filled symbols means results of spherical nuclei 
or the nuclei which have small deformation ($|\beta|<0.1$), 
open ones means those in deformed nuclei. 
Over $A=80$, the trend and relation among the results of each Skyrme parameter 
become clarified. 
The behaviour of $\bar{E}_{\rm GMR}$ in deformed nuclei diverges from 
the trend of spherical. 
The order of $\bar{E}_{\rm GMR}$ can be almost corresponded to 
the order of $K_{\infty}$ magnitude (refer to Tab.\ref{tab:gs3}). 

\begin{figure}[h]
 \begin{center}
  \includegraphics[keepaspectratio,width=5cm, angle=-90]{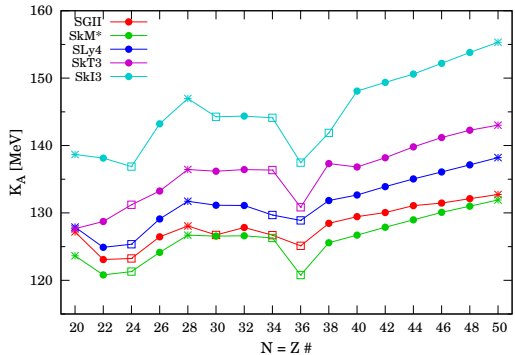}
  \caption{(Colour online) Finite nuclear incompressibility $K_A$ for 
  $N=Z$ even-even nuclide. They are estimated with Eq.(\ref{EGMR}). 
  }\label{fig:3}
 \end{center}
\end{figure}

Figure \ref{fig:3} shows  $K_A$ obtained by Eq.(\ref{EGMR}). 
Same as Fig.\ref{fig:2}, the filed and open symbols correspond to the results of 
spherical and deformed nuclei, but we use the star symbol 
for double magic (DM) nuclei ($^{40}$Ca, $^{56}$Ni, $^{100}$Sn). 
The behaviour of $K_A$ well corresponds to that of $\bar{E}_{\rm GMR}$, 
thus the trend and the relation among interactions are stable over $A=80$, 
and the results of deformed nuclei clearly have difference from spherical nuclei. 

To obtain the expansion coefficients ($K_{\rm surf}$, $K_\tau$, $K_{\rm Coul}$), 
we analyse our results according to the $K_A$ expansion in Eq.(\ref{KAex}). 
The results of deformed nucleus are excluded from our analysis, 
because they have clearly a different trend from those of spherical nuclide. 
If they can be included into the analysis, we will need the way to separate 
quadrupole and monopole modes. 
Here, two cases are considered: (i) excluding DM nuclei 
and (ii) including them. 
Our purpose is to extract nuclear matter properties from nucleus. 
The effects due to the special nuclear structure such as a modes coupling in 
deformed nuclei, should be excluded from the matter property analysis. 
As mentioned in Ref.\cite{Kha09}, the magicity effects in the incompressibility 
will appear, which should be confirmed in the comparison the (i) and (ii).

In this section, we fix the $K_{\rm surf}$ and $K_{\rm Coul}$ 
expansion coefficients which are listed in the Table \ref{tab:gs3}.
The surface term $K_{\rm surf}$ is often estimated 
as an opposite sign of $K_\infty$ \cite{SY07,Du12,Li07,PGF12,PGF13},
and the Coulomb term $K_{\rm Coul}$ is estimated in Ref.\cite{Bla80,SY07} as 
\begin{eqnarray}
\tilde{K}_{\rm Coul} = - \frac{3}{5} \frac{e^2}{R'}\left( \frac{Q_0}{K_\infty} + 8 \right), \ 
R' \equiv \left( \frac{3}{4\pi \rho_0} \right)^{1/3}\!.
\label{Kci}
\end{eqnarray}
Our $|K_{\rm surf}|$ are larger than $|K_\infty|$, 
which have been mentioned already in Ref.\cite{SY07}. 
Although an assumption $K_{\rm surf} \equiv -K_\infty$ is 
sometimes used for non-relativistic models in previous analyses \cite{SY07,Du12,Li07,PGF12,PGF13,Pat02}, 
it is not suitable actually and the difference over 30\% from it 
might cause a serious missing in nuclear property. 
In both (i) and (ii), $K_{\rm Coul}$ closes to the $\tilde{K}_{\rm Coul}$. 
$K_{\rm surf}$ and $K_{\rm Coul}$ in (ii) are a little weaker and stronger 
than those of (i), respectively in most interaction-cases. 
The effect of DM on $K_A$ is regarded as small as shown in Fig.\ref{fig:3}, 
although small kinks appear at $^{56}$Ni. 
The nuclear magicity is not so sensitive to the $K_A$ excluding light DM nuclei. 
\begin{table}[h]
\caption{Root mean square radius $\tilde{R}=\sqrt{\langle r^2 \rangle}$ [fm] and 
absolute value of quadrupole deformation parameter $\beta$ in ground state for 
$N=Z$ and $A=100, 132$ nuclide.}   
\label{tab:gs1}
\begin{tabular}[t]{ccccccccccccccc} \hline \hline
        &\multicolumn{2}{c}{SGII}&&\multicolumn{2}{c}{SkM$^*$}&&\multicolumn{2}{c}{SLy4}&&\multicolumn{2}{c}{SkT3}&&\multicolumn{2}{c}{SkI3}\\
\cline{2-3}\cline{5-6}\cline{8-9}\cline{11-12}\cline{14-15}
        &$\tilde{R}$ & $|\beta|$&&$\tilde{R}$& $|\beta|$&&$\tilde{R}$& $|\beta|$&&$\tilde{R}$& $|\beta|$&&$\tilde{R}$&$|\beta|$\\ \hline
$^{ 40}$Ca  &  3.35  &  0.00 &&  3.40  &  0.00  &&  3.39  &  0.00  &&  3.40  &  0.00  &&  3.36  &  0.00  \\
$^{ 44}$Ti  &  3.44  &  0.00  &&  3.48  &  0.00  &&  3.48  &  0.00  &&  3.47  &  0.00  &&  3.46  &  0.00  \\
$^{ 48}$Cr  &  3.56  &  0.26  &&  3.58  &  0.21  &&  3.60  &  0.26  &&  3.53  &  0.03  &&  3.58  &  0.29  \\
$^{ 52}$Fe  &  3.58  &  0.00  &&  3.61  &  0.00  &&  3.62  &  0.00  &&  3.60  &  0.00  &&  3.60  &  0.00  \\
$^{ 56}$Ni  &  3.64  &  0.00  &&  3.66  &  0.00  &&  3.67  &  0.00  &&  3.65  &  0.00  &&  3.65  &  0.00  \\
$^{ 60}$Zn  &  3.77  &  0.20  &&  3.75  &  0.00  &&  3.77  &  0.02  &&  3.74  &  0.00  &&  3.76  &  0.02  \\
$^{ 64}$Ge  &  3.83  &  0.00  &&  3.84  &  0.00  &&  3.86  &  0.00  &&  3.82  &  0.00  &&  3.86  &  0.00  \\
$^{ 68}$Se  &  3.95  &  0.22  &&  3.94  &  0.09  &&  3.97  &  0.17  &&  3.91  &  0.01  &&  3.97  &  0.23  \\
$^{ 72}$Kr  &  4.02  &  0.19  &&  4.07  &  0.26  &&  4.05  &  0.17  &&  4.03  &  0.23  &&  4.08  &  0.31  \\
$^{ 76}$Sr  &  4.06  &  0.01  &&  4.12  &  0.00  &&  4.08  &  0.01  &&  4.15  &  0.39  &&  4.08  &  0.13  \\
$^{ 80}$Zr  &  4.12  &  0.00  &&  4.15  &  0.00  &&  4.15  &  0.00  &&  4.12  &  0.00  &&  4.12  &  0.00  \\
$^{ 84}$Mo  &  4.17  &  0.00  &&  4.20  &  0.00  &&  4.20  &  0.00  &&  4.17  &  0.00  &&  4.18  &  0.00  \\
$^{ 88}$Ru  &  4.22  &  0.00  &&  4.25  &  0.00  &&  4.25  &  0.00  &&  4.23  &  0.00  &&  4.23  &  0.00  \\
$^{ 92}$Pd  &  4.27  &  0.00  &&  4.29  &  0.00  &&  4.30  &  0.00  &&  4.27  &  0.00  &&  4.28  &  0.00  \\
$^{ 96}$Cd  &  4.32  &  0.00  &&  4.34  &  0.00  &&  4.34  &  0.00  &&  4.32  &  0.00  &&  4.32  &  0.00  \\
$^{100}$Sn  &  4.36  &  0.00  &&  4.38  &  0.00  &&  4.39  &  0.00  &&  4.36  &  0.00  &&  4.36  &  0.00  \\
\hline
$^{100}$Kr  &  4.50  &  0.23  &&  4.54  &  0.23  &&  4.54  &  0.22  &&  4.50  &  0.21  &&  4.57  &  0.24  \\
$^{100}$Sr  &  4.54  &  0.39  &&  4.56  &  0.38  &&  4.57  &  0.39  &&  4.52  &  0.36  &&  4.59  &  0.40  \\
$^{100}$Zr  &  4.52  &  0.38  &&  4.52  &  0.36  &&  4.54  &  0.36  &&  4.48  &  0.33  &&  4.59  &  0.43  \\
$^{100}$Mo  &  4.40  &  0.00  &&  4.42  &  0.00  &&  4.43  &  0.00  &&  4.38  &  0.00  &&  4.46  &  0.20  \\
$^{100}$Ru  &  4.39  &  0.00  &&  4.40  &  0.00  &&  4.42  &  0.00  &&  4.37  &  0.00  &&  4.43  &  0.19  \\
$^{100}$Pd  &  4.38  &  0.00  &&  4.39  &  0.00  &&  4.40  &  0.00  &&  4.36  &  0.00  &&  4.39  &  0.01  \\
$^{100}$Cd  &  4.37  &  0.00  &&  4.38  &  0.00  &&  4.39  &  0.00  &&  4.35  &  0.00  &&  4.38  &  0.00  \\
\hline
$^{132}$Sn  &  4.78  &  0.00  &&  4.80  &  0.00  &&  4.80  &  0.00  &&  4.79  &  0.00  &&  4.82  &  0.00  \\
$^{132}$Te  &  4.79  &  0.00  &&  4.80  &  0.00  &&  4.81  &  0.00  &&  4.78  &  0.00  &&  4.82  &  0.00  \\
$^{132}$Xe  &  4.79  &  0.00  &&  4.81  &  0.00  &&  4.81  &  0.00  &&  4.78  &  0.00  &&  4.82  &  0.00  \\
$^{132}$Ba  &  4.81  &  0.15  &&  4.83  &  0.16  &&  4.84  &  0.15  &&  4.79  &  0.14  &&  4.83  &  0.16  \\
$^{132}$Ce  &  4.83  &  0.21  &&  4.85  &  0.23  &&  4.86  &  0.21  &&  4.82  &  0.23  &&  4.84  &  0.21  \\
$^{132}$Nd  &  4.93  &  0.40  &&  4.91  &  0.35  &&  4.94  &  0.38  &&  4.86  &  0.32  &&  4.94  &  0.41  \\
$^{132}$Sm  &  4.94  &  0.41  &&  4.93  &  0.38  &&  4.94  &  0.38  &&  4.89  &  0.37  &&  4.94  &  0.41  \\
\hline\hline
\end{tabular} 
\end{table}
\label{sec:NeZ}

\subsection{$A=100, 132$}
\begin{figure}[h]
 \begin{center}
  \includegraphics[keepaspectratio,width=5cm, angle=-90]{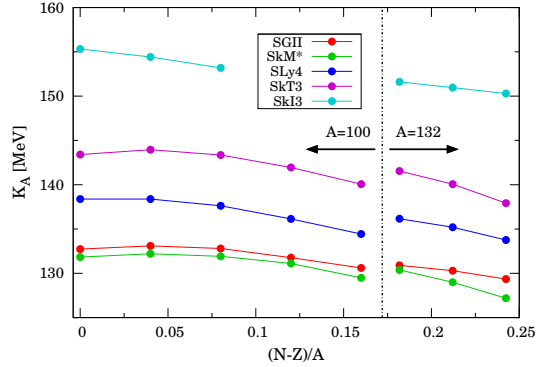}
  \caption{(Colour online) Same as Fig.\ref{fig:3}, but for 
 $K_A$ of spherical $A=100, 132$ isobar with respect to $(N-Z)/A$.
  }\label{fig:4}
 \end{center}
\end{figure}

In this section, we determine the isospin term $K_\tau$ from isobar nuclide 
$A$=100 and 132, while using $K_{\rm surf}$ and $K_{\rm Coul}$ 
fixed in previous section. 
In same as Sec. \ref{sec:NeZ}, we exclude the deformed nuclei from the analysis. 
Figure \ref{fig:4} shows the $K_A$ of the spherical isobars with $A$=100 and 132
with respect to isospin asymmetry $(N-Z)/A$, in which 
the vertical chain line separates $A$=100 and 132.
The root mean square radii and quadrupole deformations of the isobars 
are listed in Table \ref{tab:gs1}. 

The $K_A$ in Fig.\ref{fig:4} has a parabolic shape in $(N-Z)/A$ which corresponds 
to the expansion Eq.(\ref{KAex}), however the centre of the parabolic function is 
not always at $N$=$Z$. 
In this analysis, we also consider the two cases: (i) with DM and (ii) without DM. 
The isospin term $K_\tau$ obtained in the cases are listed in Table \ref{tab:gs3}. 
The effects of DM are not so large also in $K_\tau$ excluding the results of SkI3. 
When we exclude the result of DM from SkI3 results, 
the points which can be used in the analysis are only four, 
therefore the analysis ambiguity becomes large. 
To determine the $K_\tau$, the number of isobar nuclide is essential. 

In several papers, an isospin dependence of incompressibility for nuclear matter 
is estimated at saturation density $\rho_0$ 
with a small isospin asymmetry \cite{PC09,Du12,SSM14}, 
which can be written in 
\begin{eqnarray}
K_\tau^\infty = K_{\rm sym} - 6L - \frac{Q_0}{K_\infty}L . 
\label{Kti}
\end{eqnarray}
The $K_\tau^\infty$ can not be regarded as the finite incompressibility $K_\tau$, 
which is mentioned also in Ref.\cite{PGF12}, although 
the strong correlation between $K_\tau$ and $K_\tau^\infty$ can be expected naively. 
The $K_\tau^\infty$ is often used to expand the $K_\tau$: 
$K_\tau = K_\tau^\infty + K_\tau^{\rm surf}A^{-1/3}$. 
The $K_\tau^{\rm surf}$ could not be estimated 
because the number of isobar-chain sample is only two. 
The absolute value of $K_\tau$ is usual smaller than $K_\tau^\infty$ in present work 
and the correlation among them does not seem simple. 
\label{sec:isobar}

\subsection{Sn isotope}
\begin{figure}[h]
 \begin{center}
  \includegraphics[keepaspectratio,width=5cm, angle=-90]{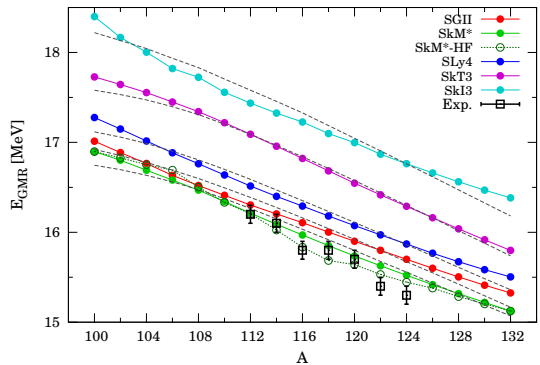}
  \caption{(Colour online) Same as Fig.\ref{fig:2}, but for 
  Sn isotopes from $A=$100 to 132 with respect to $A$.
  Filled and open circle means the result calculated by Cb-TDHFB
  and by TDHF with SkM$^*$, respectively. 
  Square symbols shows experimental data at RCNP\cite{Li07}.
  }\label{fig:5}
 \end{center}
\end{figure}

We obtain the expansion coefficients of $K_A$ in previous sections. 
To confirm the coefficients and the expansion Eq.(\ref{KAex}) itself, 
we compare the $E_{\rm GMR}$ in Eq.(\ref{EGMR}) 
with the coefficients ($K_{\rm surf}, K_\tau, K_{\rm Coul}$), in the experiment and 
$\tilde{E}_{\rm GMR}$ by the linear response calculation with Eq.(\ref{bGMR}), 
for Sn isotopes ($A$=100 - 132). 
Figure \ref{fig:5} shows the $E_{\rm GMR}$ for Sn isotopes. 
Solid lines and filled symbols mean the results directly calculated with Eq.(\ref{bGMR}), 
dashed lines correspond to the $E_{\rm GMR}$ in Eq.(\ref{EGMR}) with the coefficients 
of case (i) for each interaction, 
and open square symbols are experimental data at RCNP\cite{Li07}.
Furthermore, to show the pairing effects we add the $\tilde{E}_{\rm GMR}$ 
with TDHF only for SkM$^*$, which are symbolized by open circles and dotted line.

The dashed lines well reproduce whole $\bar{E}_{\rm GMR}$ of Sn isotopes
within a smaller than 0.3 MeV. 
It means the expansion Eq.(\ref{KAex}) is an effective procedure 
and the coefficients are suitable. 
The comparison our results and he experimental data may recommend  
SkM$^*$ and SGII parameters as a candidate of the ``{\it answer}''.

The pairing effects to the trend on Sn isotope ISGMR can be discussed 
in the comparison between our results obtained by TDHF (open) 
and by Cb-TDHFB (filled) with SkM$^*$.
The small difference between them appears in whole isotopes, 
which can be expected due to the small deformation of HF ground states. 
While the effect to soften EOS slightly in a surface-type pairing functional 
was reported \cite{Kha09,KMC10,AB13}, 
our results indicate the opposite effects, however whose mechanism is 
different from the previous studies. 
The HF ground states in Sn isotopes have some deformed aspects 
as Table \ref{tab:gs2}. 
As the explanation in Sec. \ref{sec:NeZ}, the centroid energy of ISGMR is estimated 
at lower than that of spherical nucleus due to the coupling with other modes, 
while using the summation analysis such as Eq.(\ref{bGMR}). 
\begin{table}[h]
\caption{Same as Table \ref{tab:gs1}, but for Sn isotopes with from $N=$52 to 80.
Although the HF+BCS ground states take only a spherical shape in present work, 
the HF results in SkM$^*$ have some deformation.} 
\label{tab:gs2}
\begin{tabular}[t]{cccccccccc} \hline \hline
            & SGII   &&\multicolumn{3}{c}{SkM$^*$}&&  SLy4   &  SkT3   &  SkI3   \\
\cline{4-6}\\[-4mm]
             &$\tilde{R}$&&  $\tilde{R}$  & $\tilde{R}_{\rm HF}$ & $|\beta_{\rm HF}|$&&  $\tilde{R}$ & $\tilde{R}$ & $\tilde{R}$ \\ \hline
$^{102}$Sn  &  4.39  && 4.405  &  4.401  &  0.04  &&  4.42  &  4.38  &  4.40  \\
$^{104}$Sn  &  4.43  && 4.434  &  4.426  &  0.05  &&  4.45  &  4.40  &  4.43  \\
$^{106}$Sn  &  4.46  && 4.462  &  4.448  &  0.00  &&  4.48  &  4.43  &  4.47  \\
$^{108}$Sn  &  4.48  && 4.491  &  4.488  &  0.07  &&  4.51  &  4.46  &  4.50  \\
$^{110}$Sn  &  4.51  && 4.521  &  4.527  &  0.10  &&  4.54  &  4.49  &  4.53  \\
$^{112}$Sn  &  4.54  && 4.551  &  4.558  &  0.10  &&  4.57  &  4.52  &  4.56  \\
$^{114}$Sn  &  4.57  && 4.581  &  4.596  &  0.11  &&  4.60  &  4.54  &  4.59  \\
$^{116}$Sn  &  4.59  && 4.609  &  4.632  &  0.15  &&  4.62  &  4.57  &  4.61  \\
$^{118}$Sn  &  4.62  && 4.636  &  4.699  &  0.28  &&  4.65  &  4.60  &  4.64  \\
$^{120}$Sn  &  4.64  && 4.662  &  4.677  &  0.17  &&  4.67  &  4.63  &  4.67  \\
$^{122}$Sn  &  4.67  && 4.688  &  4.703  &  0.14  &&  4.70  &  4.66  &  4.70  \\
$^{124}$Sn  &  4.69  && 4.712  &  4.724  &  0.10  &&  4.72  &  4.68  &  4.72  \\
$^{126}$Sn  &  4.71  && 4.736  &  4.743  &  0.00  &&  4.74  &  4.71  &  4.75  \\
$^{128}$Sn  &  4.74  && 4.759  &  4.766  &  0.06  &&  4.76  &  4.74  &  4.77  \\
$^{130}$Sn  &  4.76  && 4.781  &  4.784  &  0.04  &&  4.79  &  4.76  &  4.80  \\
\hline \hline
\end{tabular} 
\end{table}

\begin{table*}[b]
\caption{Parameters related to EOS ($\rho_0$[fm$^{-3}$], $K_\infty$, 
$Q_0$, $L$, $K_{\rm sym}$, $K_\tau^\infty$, $\tilde{K}_{\rm Coul}$ [MeV]) 
and finite incompressibility 
($K_{\rm surf}$, $K_\tau$, $K_{\rm Coul}$ [MeV]) for each Skyrme interaction. 
The $K_\tau^\infty$ and $\tilde{K}_{\rm Coul}$ 
are obtained by Eq.(\ref{Kti}),(\ref{Kci}).
Nucleon mass $mc^2$=938.9187 MeV, 
$\hbar c$=197.327 MeV fm and $\alpha^{-1}$=137.036 are used \cite{Cha97}.} 
\label{tab:gs3}
\begin{tabular}{cccccrcccccccccc}
\hline\hline
          &\multicolumn{7}{c}{EDF}&&\multicolumn{3}{c}{(i) without DM}&&\multicolumn{3}{c}{(ii) with DM}\\
\cline{2-8}\cline{10-12}\cline{14-16}\\[-4mm]
Int.    &$\rho_0$& $K_\infty$ & $Q_0$ & $L$ & $K_{\rm sym}$ & $K_\tau^\infty $ & $\tilde{K}_{\rm Coul}$ && $K_{\rm surf}$ & $K_\tau$ & $K_{\rm Coul}$ &&$K_{\rm surf}$ & $K_\tau$ & $K_{\rm Coul}$\\ \hline
SGII    & .1583 & 214.6 & -380.9 & 37.63 & -145.9 & -304.9 & -4.69  && -273.7 & -295.9 &-4.52 && -260.4 & -283.5 & -5.10 \\
SkM$^*$ & .1603 & 216.6 & -386.1 & 45.78 & -155.9 & -349.0 & -4.70  && -285.3 & -282.0 &-4.76 && -276.7 & -285.0 & -5.09 \\
SLy4    & .1595 & 229.9 & -363.1 & 45.96 & -119.7 & -322.9 & -4.85  && -318.4 & -314.3 &-4.73 && -308.3 & -305.4 & -5.13 \\
SkT3    & .1610 & 235.7 & -382.7 & 55.31 & -132.1 & -374.1 & -4.83  && -326.1 & -305.1 &-4.54 && -325.0 & -305.2 & -4.51 \\
SkI3    & .1598 & 258.2 & -303.8 & 100.5 &  73.03 & -411.8 & -5.13  && -372.6 & -352.0 &-4.75 && -364.8 & -332.6 & -4.99 \\
\hline\hline
\end{tabular} 
\end{table*}

\label{sec:tin-iso}

\section{Discussion}
We obtain the expansion coefficients ($K_{\rm surf}, K_\tau, K_{\rm Coul}$) 
through the Sec.\ref{sec:NeZ} and \ref{sec:isobar}, and confirm them 
in Sn isotopes. 
SkM$^*$ and SGII parameters might be likely candidates in Sec.\ref{sec:tin-iso}, 
however it is not clear that they can approach to more heavier system, 
at least in present results. 
The many GMR distributions for whole mass region have been measured in the past, 
which are summarised in Ref.\cite{SY93}. 
We extend the mass region to apply Eq.(\ref{EGMR}), 
and attempt to narrow down the parameters. 
\begin{figure}[h]
 \begin{center}
  \includegraphics[keepaspectratio,width=6cm, angle=-90]{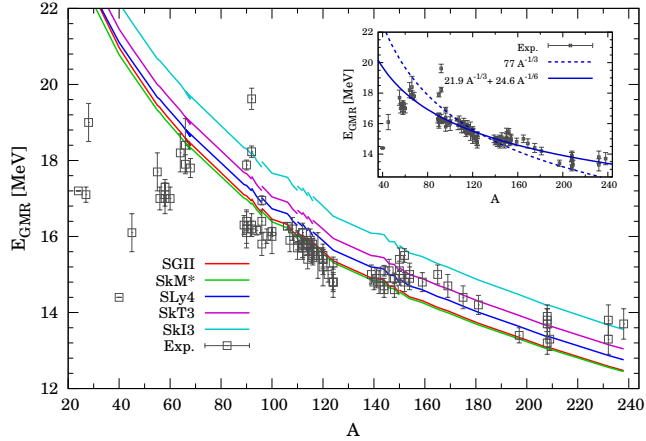}
  \caption{(Colour online) The peak positions of GMR in experiment (square)
  and the lines obtained from Eq.(\ref{EGMR}) with the $K_A$ for each Skyrme parameter, 
  where the the mean radii $\langle r^2 \rangle$ are calculated with 
  SkM$^*$ parameter, are shown. 
  }\label{fig:6}
 \end{center}
\end{figure}

Figure \ref{fig:6} shows experimental $E_{\rm GMR}$s (square) which are measured at 
Texas A\&M, Grennoble, Groningen \cite{SY93} and RCNP\cite{Li07,PGF12}, 
and the centroid energy (solid line) by Eq.(\ref{EGMR}) for each Skyrme parameter  
but with $\langle r^2 \rangle$ expressed as 
$\sqrt{\langle r^2 \rangle}$= 0.895$A^{1/3}+$ 0.321 fm 
which is obtained by fitting systematic mean radii calculated in SkM$^*$.
The experimental data cover a wide mass region $A$=24$-$238 with $Z$=12$-$92. 
In $A<$50 mass region, 
the mass dependence of experiment values is not clear apparently. 
Over $A$=100, they make a visible trend, though some values around $A$=90 
deviate from the trend. 

Whole mass region, the solid lines keep their magnitude order 
which is same one in Fig.\ref{fig:5}, thus 
namely SkI3, SkT3, SLy4, SGII and SkM$^*$ in decreasing order.  
For SkI3, the results overestimate experiments. 
For SGII and SkM$^*$ they underestimate over $A$=160, 
while they can well reproduce those around $A$=110. 
If we emphasize the agreement with a heavy system, 
SGII and SkM$^*$ can not reach to the conclusion as the best choice. 
SkI3 should be excluded from the candidate, due to the missing over $A$=100. 
From the agreements with $E_{\rm GMR}$ over $A$=140, 
SLy4 and SkT3 are good candidates. 
Therefore our results indicate that the range of $K_\tau$ is -305$\pm$10 MeV, 
and the incompressibility parameters for infinite system 
corresponding to the $K_\tau$ are $K_\tau^\infty$=-340$\pm$35, 
$K_\infty$=225$\pm$11 and $K_{\rm sym}$=-138$\pm$18 MeV, 
which are expected from SGII, SkM$^*$, SLy4 and SkT3 parameters. 

In this paper, our analysis has constructed within
the assumption Eq.(\ref{EGMR}) and (\ref{KAex}). 
As Fig.\ref{fig:6} shown that, the expansion Eq.(\ref{KAex}) 
well work to connect between the infinite nuclear properties and 
finite nuclear excitation mode in Eq.(\ref{EGMR}). 
However, the mass dependence of $E_{\rm GMR}$ in experiment 
is apparently different from that scaled by $A^{-1/3}$, 
which has been mentioned also in Ref.\cite{AB13,CP14,Pie10}.
The small panel in Fig.\ref{fig:6} shows 
the fitting experimental data for $A$=90$-$238, 
by $A^{-1/3}$ (dashed) and by $A^{-1/3}+A^{-1/6}$ (solid).
The coefficient of $A^{-1/3}$ is consistent 
with the ordinary value \cite{Liu91}, but it underestimates experiment 
in heavy mass region. 
The solid line reproduces experimental data in a whole mass region.
The $A^{-1/3}$ and $A^{-1/6}$ scaling are well known on the GDR 
as Steinwedel-Jensen (SJ) and Goldhaber-Teller (GT) model \cite{GM95}, 
respectively 
in which the dipole moments are caused by polarization and 
by proton-neutron displacement.
If we adapt the models to GMR, 
the SJ and GT-model will respectively correspond to 
the density wave oscillation keeping nuclear size, 
and to the vibration between expansion and contraction of the size. 
\label{sec:discuss}

\section{Conclusion}
We have performed a systematic investigation of the ISGMR to 
extract the isospin dependent compression EOS parameter from 
finite nuclear system, using the linear response calculation with 
Cb-TDHFB represented in 3D coordinate space. 
The expansion coefficients ($K_{\rm surf}$, $K_\tau$, $K_{\rm Coul}$) 
of finite incompressibility $K_A$ are determined from 
$N$=$Z$ and isobar $A$=100, 132 nuclide for each Skyrme interaction
(SGII, SkM$^*$, SLy4, SkT3, SkI3).
Furthermore in Sn isotopes ($A$=100$-$132), it is confirmed that 
the coefficients are suitable and the $K_A$ expansion is available.

The absolute values of $K_{\rm surf}$ are lager than those of $K_\infty$,
thus the analysis assuming $K_{\rm surf}$=$-K_\infty$ is not effective, 
which are often used in previous study for non-relativistic interaction.
The $K_{\rm Coul}$ estimation Eq.(\ref{Kci}) works well. 
The isospin term $K_\tau$ in the finite system 
is smaller than the $K_\tau^\infty$. 
The magicity in a present work does not affect the conclusion. 
And the pairing effect in Sn isotopes are not so large, 
although the deformation due to the lack of pairing causes 
the coupling quadrupole and monopole modes, 
and disturbs the position of GMR.

We narrow down the values from the comparison experimental data 
with our results in a whole mass region, 
which indicates the range of $K_\tau$ is -305$\pm$10 MeV.
The incompressibility parameters corresponding to the range 
have $K_\infty$=225$\pm$11 and $K_{\rm sym}$=-138$\pm$18 MeV, 
which is consistent with the previous study of slop parameter $L$ \cite{Br13}.
As an indication of Eq.(\ref{Kti}), 
the $L$ has an important role to decide $K_{\rm sym}$. 
Thus the studies related to $L$ should be also pressed forward in parallel. 
The ordinary mass scaling of GMR has underestimation 
in heavy mass system, which may indicate the needs to reconsider 
the expression of incompressibility.

To determine the isospin term $K_\tau$, the systematic ISGMR data 
in isobar nuclide are very effective. 
They should be heavy system over $A$=100 at least, 
because in light system the nuclear speciality is showing up. 
When the heavy nucleus is used to analyse GMR, 
we should note the deformation of open shell nuclei. 
If the monopole and other modes in deformed nucleus can be separated, 
the analysis data increases and 
the relation EOS parameter and finite system will be more robust.
\label{sec:con}

\section*{Acknowledgments} 
The author greatly appreciates Professor M. Kimura
for discussions and encouragements. 
Computational resources were mostly provided by the 
High Performance Computing system 
at Research Center for Nuclear Physics, Osaka University.


\end{document}